\documentclass[twocolumn]{aastex631}

\usepackage{amsmath}
\usepackage{color}

\newcommand{\kms}{km\,s$^{-1}$}
\newcommand{\bl}[1]{#1}
\newcommand{\rd}[1]{#1}
\newcommand{\gn}[1]{#1}
\newcommand{\mg}[1]{#1}
\newcommand{\rv}[1]{#1}
\newcommand{\rvb}[1]{#1}

\makeatletter
\def\endacknowledgments{\expandafter\ifx\csname internallinenumbers\endcsname\relax\else\end{internallinenumbers}\fi
\egroup\unvbox\ackbox\vskip6pt}
\makeatother

\shorttitle{Suprathermal CH$^+$ rotational emissions in HH 168}
\shortauthors{Neufeld et al.}

\hypersetup{hidelinks}

\begin{document}

\title{First detection of suprathermal CH$^+$ rotational emissions from chemical pumping:\\ JWST/MIRI observations of the HH 168 molecular shock}

\author[0000-0001-8341-1646]{David A. Neufeld}
\affiliation{William H. Miller III Department of Physics and Astronomy, Johns Hopkins University, 3400 North Charles Street, Baltimore, MD 21218, USA}
\email{neufeld@jhu.edu}

\author[0000-0003-1665-5709]{Joel D. Green}
\affiliation{Space Telescope Science Institute, 3700 San Martin Drive, Baltimore, MD 21218, USA}

\author[0000-0001-8302-0530]{Dan M. Watson}
\affiliation{Department of Physics and Astronomy, University of Rochester, Rochester, NY 14627, USA}

\author[0000-0003-2027-5020]{Paule Sonnentrucker}
\affiliation{European Space Agency (ESA) Office, Space Telescope Science Institute, 3700 San Martin Drive, Baltimore, MD 21218, USA}

\begin{abstract}
The methylidyne cation forms in warm or shocked \mg{interstellar} gas through the endothermic reaction C$^+$ + H$_2$ $\to$
CH$^+$ + H. \rd{Because C$^+$, which is common in irradiated and shocked gas, reacts with H$_2$ only when the reactants supply $\gtrsim 0.4$\,eV, CH$^+$ traces energy deposition in molecular gas.} \mg{JWST/MIRI observations of HH 168} reveal the pure rotational lines \bl{R(17)}--R(31) of
CH$^+$ at \rd{13.7--21.3}\,$\mu$m, from upper levels $J = \bl{18}$--32 \mg{(}\bl{0.57}--1.65\,eV\mg{)}. Eleven lines exceed 5\,$\sigma$ in \rv{both of the two fields that we observed} and match the hot H$_2$ \rd{in \mg{radial} velocity and
spatial distribution}. Because \mg{collisional excitation is} negligible, each ion entering the $v = 0$
ladder emits one photon in every R($J$) line below its entry level, so the line intensities \mg{provide the first measured rotational distribution of the ground-state products}. The photon intensity falls almost linearly with \mg{increasing} $J$, as expected \mg{for formation} at a similar rate into every level up to $J \approx 33$; formation in $J = 32$ requires \mg{$\geq 2$\,eV} of reactant energy. The line positions\mg{,} up to 150\,\kms\ redward of ExoMol predictions, provide the first measurement of
the CH$^+$ rotational intervals above $J = 16$. \mg{A Boltzmann distribution at $\phi T \approx 6200$\,K, where $\phi$ is the fraction of the available energy appearing in rotation, fits slightly better, implying reactant energies equivalent to several thousand kelvin.} These lines \mg{probe} the dynamics of a prototypical \mg{interstellar} ion-molecule reaction.
\end{abstract}


\section{Introduction} \label{sec:intro}

The methylidyne cation was among the first interstellar molecules to be identified \citep{Douglas1941}, and its
abundance in diffuse clouds has posed a long-standing puzzle, because its only efficient formation route,
\begin{equation}
{\rm C^+ + H_2 \to CH^+ + H}, \qquad \Delta E = +0.40\,{\rm eV},
\label{eq:form}
\end{equation}
is endothermic by 0.40\,eV and cannot proceed in cold gas \citep{Elitzur1978}. The proposed solutions, shocks
\citep{Elitzur1978,Draine1986,Pineau1986}, turbulent dissipation \citep{Godard2009,Falgarone2010} and, in dense photodissociation regions,
vibrationally excited H$_2$ \citep{Agundez2010,Nagy2013}, all supply the energy
of reaction (\ref{eq:form}) from the relative motion of C$^+$ and H$_2$ or from the internal energy of H$_2$.

Pure rotational emission of CH$^+$ was first detected with ISO in the planetary nebula NGC 7027
\citep{Cernicharo1997} and has since been observed with \rd{the far-infrared Herschel Space Observatory} in the Orion Bar, in lines up to $J = 6 \to 5$ \citep{Nagy2013} and in maps of the $J = 3 \to 2$ line \citep{Parikka2017}. The excitation of those lines already required ``chemical
pumping'': because the reaction deposits CH$^+$ in excited levels and the ion is destroyed by H\rv{,} H$_2$ \rv{and electrons} before it can
thermalize, the level populations \mg{of the observed molecules} retain a memory of the formation process \citep{Godard2013,Faure2017}. The same
chemical pumping accounts for the vibrational emission of CH$^+$ in NGC 7027, where the $v = 1 \to 0$
band observed in the near infrared arises from ions formed vibrationally excited by reaction (\ref{eq:form}) with
vibrationally excited H$_2$ (\citealt{Neufeld2021}; see also \citealt{Changala2021}, and \citealt{Sil2026} for the effect of chemical pumping on the level populations). A close analogue is the
suprathermal (``prompt'') emission of OH from rotational levels up to $N = 44$--46, which traces the nascent distribution of the
OH produced by photodissociation of H$_2$O \citep{Tappe2008,Tabone2021,Zannese2024,Neufeld2024}; there the excess excitation
comes from a photodissociation, here from a chemical reaction.

The product state distribution of reaction (\ref{eq:form}) itself has never been measured for the ground
electronic state of CH$^+$. Chemiluminescence experiments have determined the rotational and vibrational distributions of
the electronically excited A\,$^1\Pi$ product at collision energies of 3--8\,eV \citep{Kusunoki1979}, and quantum
calculations of state-to-state rate coefficients exist for the ground state at thermal energies \citetext{\citealt{Zanchet2013,Faure2017}; \citealt{GonzalezLezana2026}},
but not at the collision energies of 1--3\,eV found in shocks.

Here we report JWST/MIRI observations of the Herbig-Haro object HH 168 in which the pure rotational lines of CH$^+$ are
detected from levels as high as $J = 32$, more than 19,000\,K above the ground state. The emission is suprathermal. With radiative
decay rates of 40--180\,s$^{-1}$, collisions at the densities of the shocked gas cannot maintain significant populations in
these levels, which are instead fed by the formation of CH$^+$ in highly excited states. We show that these lines
are the radiative cascade of the newly formed ions and that their intensities therefore measure the nascent rotational
distribution of the reaction products\bl{, which is well described by a Boltzmann distribution at a formation temperature of $\approx 6200$\,K}.

\section{Observations and data reduction} \label{sec:obs}

The bright Herbig-Haro object HH 168 lies in Cepheus A West, $\approx 700$\,pc away \citep{Moscadelli2009}.
Our spectra come from JWST GO program 7084 (PI: J.~D.~Green; ``Measuring the Thickness of a J-Shock in Cep A West''), which
used the MIRI Medium Resolution Spectrometer \citep[MRS;][]{Wells2015,Argyriou2023} to cover the full 4.9--27.9\,$\mu$m
range with resolving power $R \approx 1300$--3700. On 2026 August 1, observations 1 and 2 mapped two fields, G and C \rd{(named after the shock knots in Figure~\mg{5} of \citealt{Hartigan2000})},
along the shock, each with a five-pointing mosaic and the 4-point dither in every grating setting. Exposures used FASTR1
readout with 57 groups and two integrations per exposure (319\,s), giving 6383\,s per setting, and observation 3 provided a
separate background field\rd{, at a single pointing,} taken with identical settings.
We processed the calibrated exposures retrieved from MAST (pipeline version 2.0.1) through stage 3 with JWST pipeline version 3.0.0 and Calibration Reference Data System context {\tt jwst\_1535.pmap}, subtracting the background
exposures at the calibrated-exposure stage, ahead of cube building. The data can be retrieved from MAST at \dataset[doi:10.17909/6wh2-6424]{https://doi.org/10.17909/6wh2-6424}.

All line fluxes are measured in maps brought to the angular resolution of the MRS at $23\,\mu$m ($0.87''$ FWHM). For each
line we fitted, spaxel by spaxel, a Gaussian whose center and width were fixed (the latter at the width measured in each MIRI sub-band from free fits to the brightest unblended lines, corresponding to resolving powers of 3000--3200 in channel 3 and 2100--2300 in channel 4, higher in channel 4 than the in-flight values of \citealt{Argyriou2023}, as also found by \citealt{Pontoppidan2024b}), plus a
quadratic baseline, with neighboring lines included as additional components where necessary (Section~\ref{sec:results});
for the bright H$_2$ $v = 0$ lines the center and width were free.

Each line map was convolved with a Gaussian whose FWHM is
the quadrature difference between the $23\,\mu$m beam and the beam at the line wavelength, regridded to a common $0.35''$
grid, and summed over a contiguous region of 3.06\,arcsec$^2$ chosen, in each field, to have the highest convolved H$_2$ S(7)
intensity \rd{(outlined \gn{on the S(6) maps of} Figure~\ref{fig:maps})}. All lines are therefore measured in the same effective beam, and no aperture correction \mg{need be} applied. Uncertainties
follow from the residuals of the same fit to the aperture spectrum of the convolved cube.

Extinction corrections, $A_V = 3.0$ (G) and 4.1\,mag (C), were derived from the H$_2$ rotational lines, fitted with two temperature components as in \citet{Neufeld2024}, with the KP5
extinction curve \citep{Chapman2009,Pontoppidan2024}, and applied to all the CH$^+$ line intensities below.

\section{Results} \label{sec:results}

\subsection{Detection and identification} \label{sec:detect}

Among the \mg{hitherto} unidentified emission lines in the HH 168 spectra, a family of lines at 20.28, 19.41, 18.64, 17.31, 16.74, 16.23
and 15.76\,$\mu$m stood out by their common spatial distribution, which follows the hot H$_2$ emission. Their wavenumbers
form a smooth sequence with spacings decreasing from 22.0 to 18.4\,cm$^{-1}$, \gn{as expected for the R branch of a diatomic rotor. Spacings this large require a rotational constant $B \gtrsim 11$\,cm$^{-1}$ and hence, for any plausible bond length, a reduced mass below 1.5\,amu: the carrier had to be a hydride. The \mg{uniform and} rapid decrease of the spacings pointed to strong centrifugal distortion, as expected for \mg{a rotor this light} at high $J$.} A fit of a rotational
series $\nu = 2B(J+1) - 4D(J+1)^3 + 6H(J+1)^5$ to these lines gives $B = 13.93$\,cm$^{-1}$ and reproduces the
laboratory $J = 1 \to 0$ transition of CH$^+$ at 27.857\,cm$^{-1}$ \citep{Amano2010,Domenech2018} to within 0.006\,cm$^{-1}$.
We know of no other diatomic or light polyatomic species with a compatible rotational constant and distortion.
The identification was confirmed by the subsequent detection of the predicted neighboring members of the ladder\gn{, and the full set of line positions is analyzed in Section~\ref{sec:positions}}.

\begin{figure*}[t]
\plotone{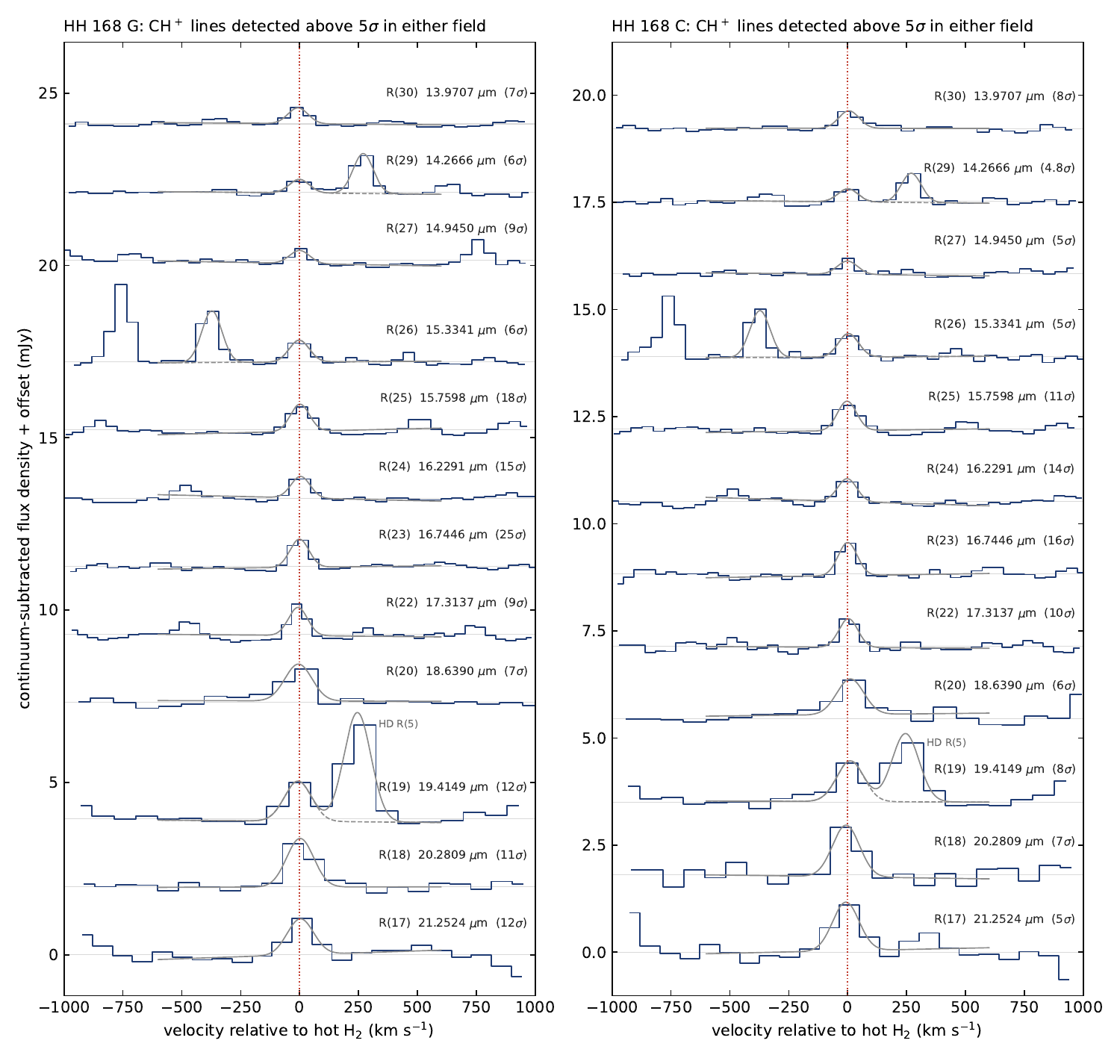}
\caption{Continuum-subtracted spectra of the CH$^+$ R($J$) lines detected at more than $5\,\sigma$ \gn{in at least one field}, in fields G (left) and C
(right), offset vertically, against velocity relative to the mean velocity of the hot H$_2$ lines S(5)--S(8) in each
field, with the rest wavelengths of Table~\ref{tab:lines}. Gray: Gaussian fits of fixed instrumental width and free center, with a linear baseline (the fluxes of Table~\ref{tab:lines} are from fixed-center fits); for R(19) the HD
R(5) line at 19.431\,$\mu$m, and for R(26) and R(29) the neighboring lines of Table~\ref{tab:lines} (notes d and e), are fitted as
additional components (dashed: the CH$^+$ component alone). \bl{R(21), at 17.94\,$\mu$m, is omitted because it lies under [Fe\,{\sc ii}] 17.936\,$\mu$m (Table~\ref{tab:lines}).}
\label{fig:lines}}
\end{figure*}

Table~\ref{tab:lines} lists the R($J$) lines of CH$^+$ from $J_{\rm up} = 17$ to 32, with their fluxes in the two fields,
and Figure~\ref{fig:lines} shows the spectra of \gn{the twelve lines detected at more than 5\,$\sigma$ in at least one field (eleven of them in both)}, with the R(23) line at 16.745\,$\mu$m the most significant (25 and 16\,$\sigma$). R(16) is \bl{only tentative ($\approx 2\,\sigma$ in each field)}, and the ladder is continuous from $J_{\rm up} = \bl{18}$ to 32 except
for R(21) at 17.94\,$\mu$m, which is hidden by [Fe\,{\sc ii}] 17.936\,$\mu$m; the next line, R(32) at 13.45\,$\mu$m ($J_{\rm up} = 33$), is
blended with a stronger unrelated line $\approx 75$\,\kms\ to its red; the lines \rd{R(0)--R(12), at $\lambda > 28\,\mu$m,} fall outside the MRS range, and \rd{R(13) and R(14), at 26.6 and 25.0\,$\mu$m} where the spectra are noisier, are not detected ($< 3\,\sigma$\rd{; upper limits in Table~\ref{tab:lines}}); R(15), at $23.59\,\mu$m, is blended with an unrelated line of \mg{similar} strength.
Einstein $A$ coefficients and level energies are taken from the ExoMol PYT line list \citep{Pearce2024}; the $A$ values,
40--180\,s$^{-1}$, are large because of the large dipole moment of CH$^+$ (1.68\,D; \citealt{Cheng2007})\rd{, eight to eleven orders of magnitude larger than those of the H$_2$ rotational lines}\mg{,}
and are insensitive to the small \mg{uncertainties} in the computed level energies discussed below.

\begin{figure*}[t]
\plotone{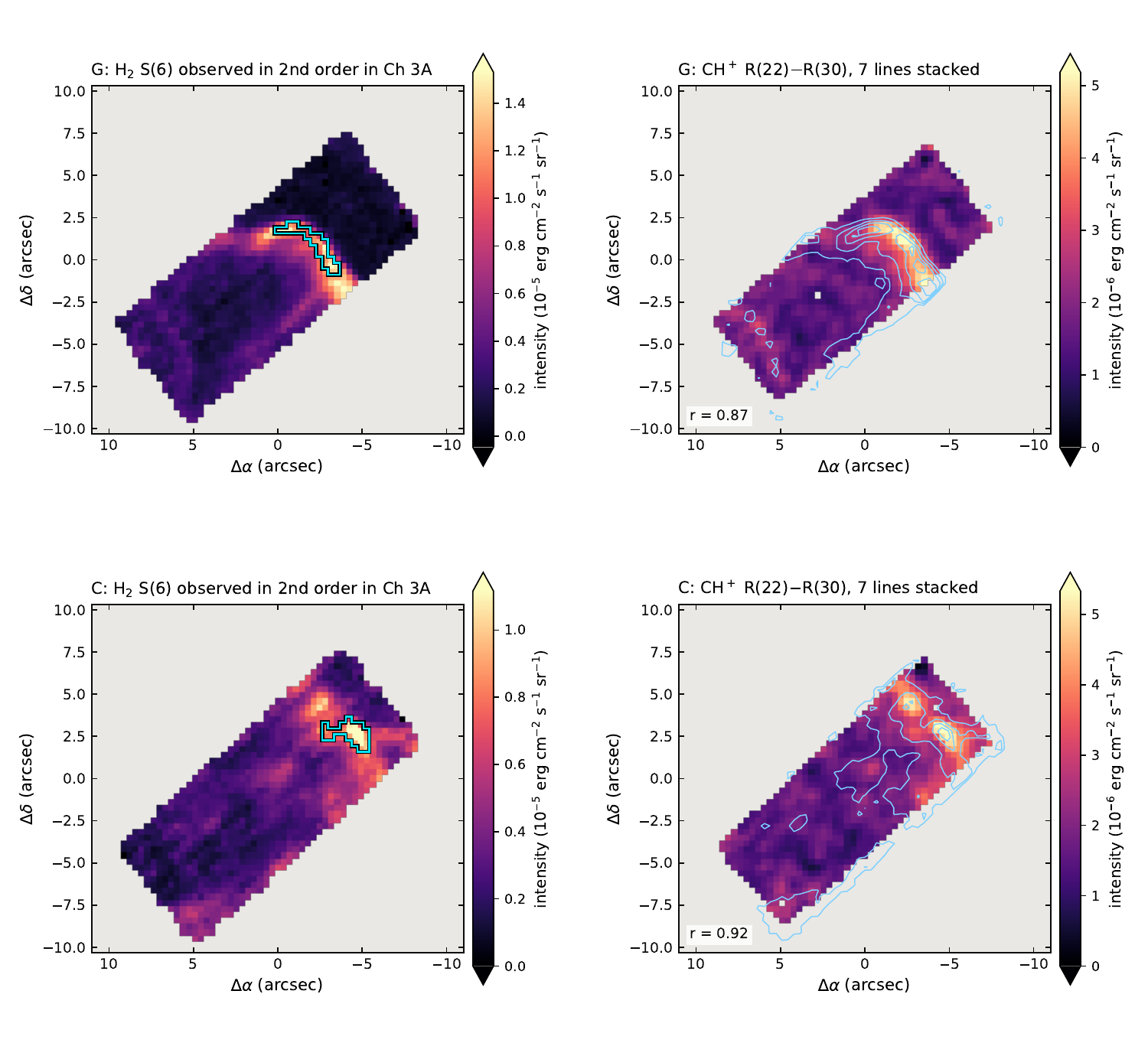}
\caption{Maps of H$_2$ S(6) observed in second order in MIRI channel 3A (left; this image carries 1.5\% of the
first-order S(6) flux, \citealt{Neufeld2026}, and is shown without further convolution, since its native resolution, $\approx 0.86''$, already matches the $0.87''$ of the CH$^+$ maps) and of the sum of seven CH$^+$ lines in channel 3, R(22), R(23),
R(24), R(25), R(27), R(28) and R(30) (right), for fields G (top) and C (bottom), at a common angular resolution.
\rd{The outlines on the S(6) maps mark the 3.06\,arcsec$^2$ apertures\gn{, chosen on the convolved S(7) maps (Section~\ref{sec:obs}),} from which the line fluxes were extracted.}
The contours drawn over the CH$^+$ maps mark 20\%, 40\%, 60\% and 80\% of the S(6) peak, and $r$ is the correlation
coefficient between the stacked CH$^+$ and the first-order S(6) map from channel 1 (not shown), computed pixel by pixel wherever the stacked CH$^+$ exceeds $3\,\sigma$ within the channel 1 field. Offsets are relative to
$\alpha = 22^{\rm h}56^{\rm m}04\fs73$, $\delta = +62\arcdeg02\arcmin15\farcs6$ (G) and
$22^{\rm h}56^{\rm m}02\fs83$, $+62\arcdeg02\arcmin10\farcs5$ (C) (J2000).
\label{fig:maps}}
\end{figure*}

The CH$^+$ lines are at the velocity of the H$_2$ emission. Relative to the mean velocity of the hot H$_2$ lines
S(5)--S(8), the line centroids measured in the two fields agree within their uncertainties (Table~\ref{tab:lines}); the H$_2$ lines S(1) and S(2), which fall in channel 3 with most of the CH$^+$ lines, are 5\,\kms\ (G) and 10\,\kms\ (C) redder than S(5)--S(8), a measure of the uncertainty in the MIRI wavelength calibration between channels, and referring the offsets to them instead would shift them by those amounts, with the two fields still agreeing ($\Delta v_{\rm G-C} = +3 \pm 2$\,\kms).

The velocity-integrated line maps (Figure~\ref{fig:maps}), stacked over seven lines that fall in MIRI channel 3,
follow the H$_2$ S(6) emission with a pixel-by-pixel correlation coefficient of 0.87 (G) and 0.92 (C). The S(6) map shown
in Figure~\ref{fig:maps} is the second-order image of the line that a spectral leak admits into channel 3A, which covers the
same field as the CH$^+$ lines; the correlation coefficients are computed instead with the first-order S(6) map from channel 1,
over the part of the field that it covers (the second-order image gives 0.74 and 0.80).

{\renewcommand{\arraystretch}{0.8}\renewcommand{\tablenotemark}[1]{{\normalfont\textsuperscript{\itshape #1}}}\renewcommand{\tablenotetext}[2]{\vskip1pt{\currtabletypesize\vskip1pt\indent\vrule height 8pt depth 2pt width0pt\relax\textsuperscript{\itshape #1}#2\vskip1pt}}\tabletypesize{\doublespacefalse\manufalse\footnotesize}
\begin{deluxetable*}{lcccccccc}
\tablecaption{CH$^+$ R($J$) lines in HH 168\label{tab:lines}}
\tablehead{\colhead{Transition} & \colhead{$\lambda$ ($\mu$m)\tablenotemark{a}} & \colhead{$A$ (s$^{-1}$)} & \colhead{$E_u/k$ (K)} & \colhead{$F_{\rm G}$} & \colhead{$F_{\rm C}$} & \colhead{$\Delta v_{\rm G}$\tablenotemark{b}} & \colhead{$\Delta v_{\rm C}$\tablenotemark{b}}}
\startdata
\rd{R(13)\tablenotemark{f}} & \rd{26.6491} & \rd{23} & \rd{4123} & \rd{$<26$} & \rd{$<38$} & \nodata & \nodata \\
\rd{R(14)\tablenotemark{f}} & \rd{25.0427} & \rd{28} & \rd{4698} & \rd{$<10$} & \rd{$<12$} & \nodata & \nodata \\
\rd{R(15)\tablenotemark{g}} & \rd{23.5997} & \rd{33} & \rd{5308} & \nodata & \nodata & \nodata & \nodata \\
R(16) & \bl{22.3484}\tablenotemark{c} & 39 & 5952 & \bl{4.9 $\pm$ 2.6} & \bl{7.0 $\pm$ 3.0} & \nodata & \nodata \\
R(17) & 21.2524 & 46 & 6629 & 7.8 $\pm$ 0.7 & 7.2 $\pm$ 1.4 & $+8 \pm 8$ & $-4 \pm 12$ \\
R(18) & 20.2809 & 53 & 7339 & 10.0 $\pm$ 0.9 & 9.3 $\pm$ 1.3 & $+20 \pm 5$ & $+11 \pm 8$ \\
R(19) & 19.4149 & 60 & 8080 & 9.0 $\pm$ 0.7 & 6.2 $\pm$ 0.7 & $+24 \pm 7$ & $+41 \pm 9$ \\
R(20) & 18.6390 & 68 & 8852 & 7.3 $\pm$ 1.0 & 5.7 $\pm$ 0.9 & $+28 \pm 8$ & $+45 \pm 11$ \\
\rd{R(21)\tablenotemark{g}} & \rd{17.9394} & \rd{76} & \rd{9653} & \nodata & \nodata & \nodata & \nodata \\
R(22) & 17.3137 & 85 & 10485 & 6.0 $\pm$ 0.7 & 4.3 $\pm$ 0.4 & $+51 \pm 6$ & $+61 \pm 4$ \\
R(23) & 16.7446 & 95 & 11344 & 5.6 $\pm$ 0.2 & 4.9 $\pm$ 0.3 & $+59 \pm 3$ & $+60 \pm 2$ \\
R(24) & 16.2291 & 104 & 12231 & 4.7 $\pm$ 0.3 & 3.9 $\pm$ 0.3 & $+74 \pm 6$ & $+68 \pm 6$ \\
R(25) & 15.7598 & 114 & 13144 & 5.5 $\pm$ 0.4 & 4.8 $\pm$ 0.5 & $+68 \pm 6$ & $+64 \pm 6$ \\
R(26)\tablenotemark{d} &15.3341 & 124 & 14083 & 4.9 $\pm$ 0.9 & 4.3 $\pm$ 0.8 & $+90 \pm 8$ & $+94 \pm 10$ \\
R(27) & 14.9450 & 134 & 15046 & 3.0 $\pm$ 0.3 & 2.2 $\pm$ 0.4 & $+97 \pm 5$ & $+95 \pm 7$ \\
R(28)\tablenotemark{e} & 14.5900 & 145 & 16032 & 2.2 $\pm$ 0.7 & 2.0 $\pm$ 0.5 & $+105 \pm 12$ & $+102 \pm 8$ \\
R(29)\tablenotemark{e} & 14.2666 & 155 & 17041 & 3.1 $\pm$ 0.5 & 2.9 $\pm$ 0.6 & $+123 \pm 6$ & $+125 \pm 10$ \\
R(30) & 13.9707 & 166 & 18072 & 3.7 $\pm$ 0.5 & 3.0 $\pm$ 0.4 & $+127 \pm 6$ & $+138 \pm 5$ \\
R(31)\tablenotemark{e} & 13.7006 & 176 & 19122 & 1.7 $\pm$ 0.4 & 1.2 $\pm$ 0.4 & $+133 \pm 18$ & $+162 \pm 18$ \\
\rd{R(32)\tablenotemark{g}} & \rd{13.455} & \rd{187} & \rd{20191} & \nodata & \nodata & \nodata & \nodata \\
\rd{R(33)\tablenotemark{h}} & \rd{13.229} & \rd{197} & \rd{21281} & \rd{$<2.3$} & \rd{$<1.6$} & \nodata & \nodata \\
\enddata
\tablenotetext{a}{Measured rest wavelength (vacuum) in the frame of the hot H$_2$ emission, from the weighted mean of the two
fields; statistical uncertainties 0.0001--0.0005\,$\mu$m (2--7\,\kms), or 0.0006\,$\mu$m (12\,\kms) for the weak R(31), to which the MIRI wavelength calibration adds
$\sim 10$\,\kms\ systematically. Measured positions are $+5$ (R(17)), $+18$ (R(18)), $+30$, $+33$, $+58$, $+60$ (R(23)), $+71$, $+66$, $+92$,
$+96$ (R(27)), $+103$, $+124$, $+134$ and $+148$\,\kms\ (R(31)) to the red of the ExoMol PYT positions \citep{Pearce2024}.}
\tablenotetext{b}{Offset of the line centroid from the ExoMol PYT position in each field, in \kms, relative to the mean
velocity of the hot H$_2$ lines S(5)--S(8) of that field; statistical uncertainties only (the MIRI wavelength calibration adds
$\sim 10$\,\kms\ between channels).}
\tablenotetext{c}{Position predicted by the rotational fit of Section~\ref{sec:detect}, at which the flux is measured; this line is
too weak for a centroid measurement. This position, close to the peak of the emission, lies \bl{0.08\,cm$^{-1}$ (54\,\kms)} to the red of that calculated from the laboratory constants of \citet{Hakalla2006}, whose data extend to $J = 17$. The ExoMol position differs from Hakalla's by $+0.6$\,cm$^{-1}$, reflecting
the transition between empirical (MARVEL) and calculated levels at $J = 16$--17.}
\tablenotetext{d}{OH 15.314\,$\mu$m and a second line $\approx 350$\,\kms\ to the blue of R(26) are fitted as one additional component.}
\tablenotetext{e}{A line $\approx 260$\,\kms\ to the red of R(29) and of R(31), and an unidentified line $\approx 175$\,\kms\ to the red of R(28), are fitted as additional components.}
\tablenotetext{f}{\rd{ExoMol PYT position; $3\,\sigma$ upper limits on the flux.}}
\tablenotetext{g}{\rd{Blended (see below); no flux is given. R(15) and R(21) at their ExoMol positions, R(32) from the fit of Section~\ref{sec:positions}.}}
\tablenotetext{h}{\rd{Position extrapolated from the trend of the offsets with $J$ (Table note a); $3\,\sigma$ upper limits.}}
\tablecomments{$A$ values and upper-level energies from the ExoMol PYT line list \citep{Pearce2024}. $F_{\rm G}$, $F_{\rm C}$:
observed (not extinction-corrected) line fluxes integrated over the $\approx 3\,\rm arcsec^2$ apertures of
Section~\ref{sec:obs}, in units of $10^{-17}\,\rm erg\,cm^{-2}\,s^{-1}$, from fits of fixed center and width (the width measured in each sub-band; statistical
uncertainties). R(15) at 23.59\,$\mu$m is blended with an unrelated line of \mg{similar} strength, R(21) at 17.94\,$\mu$m lies under [Fe\,{\sc ii}] 17.936\,$\mu$m and R(32) at 13.455\,$\mu$m (from the fit of Section~\ref{sec:positions}) is blended with
a stronger line $\approx 75$\,\kms\ to its red; \rd{these three are listed without fluxes}.}
\end{deluxetable*}}

\subsection{Line positions and the CH$^+$ energy levels} \label{sec:positions}

The measured line positions lie systematically to the red of the ExoMol positions, by an amount that grows
with $J$: $+5 \pm 7$\,\kms\ for R(17), $+18 \pm 4$ for R(18), $+60 \pm 2$ for R(23), $+96 \pm 4$ for R(27), $+134 \pm 4$ for R(30) and $+148 \pm 13$\,\kms\ for R(31)
(Table~\ref{tab:lines}), i.e., $\approx 10$\,\kms\ per unit $J$ above $J \approx 20$. The H$_2$ and HD lines in the same spectra
show no such offsets, so the shifts are not kinematic. The ExoMol energies of the CH$^+$ levels above $J = 16$ are not
constrained by laboratory data; they are computed from a potential energy curve and carry quoted uncertainties of
3--11\,cm$^{-1}$ for the observed levels \citep{Pearce2024}, whereas the cumulative offset we measure corresponds to a level error of only
$\approx 2$\,cm$^{-1}$ at $J = 31$ (relative to $J = 17$).

The only high-resolution laboratory measurements involving these levels are the A--X
photofragment lines of \citet{Sarre1986}, from which the combination differences $E(v{=}1, J{=}35) - E(v{=}1, J{=}33)$ and
$E(1,34) - E(1,32)$ can be formed; they are smaller than the ExoMol values by 0.91 and 0.77\,cm$^{-1}$, i.e., the true
rotational intervals near $J = 34$ in $v = 1$ are 160--190\,\kms\ short of the ExoMol predictions, consistent in sign and
magnitude with the trend of our $v = 0$ offsets. The wavelengths in Table~\ref{tab:lines} therefore provide the first
measurement of the rotational intervals of CH$^+$ X\,$^1\Sigma^+$ above $J = 16$, with uncertainties of
0.01--0.03\,cm$^{-1}$ for the strongest lines (including the calibration systematic). Referencing each line instead to the H$_2$ or HD lines in its own MIRI sub-band,
where such lines are available, changes the offsets by at most $\sim 10$\,\kms, \rd{comparable to the uncertainty of the MIRI wavelength calibration and} much less than the trend with $J$.

A fit of $\nu = 2BJ - 4DJ^3 + 6HJ^5 - 8LJ^7$ ($J = J_{\rm up}$; level energies $BX - DX^2 + HX^3 - LX^4$ with $X = J(J+1)$) to the 14 measured
positions, with the calibration systematic of 10\,\kms\ added in quadrature to the statistical errors, gives $B = 13.936 \pm 0.004$,
$D = (1.392 \pm 0.011) \times 10^{-3}$, $H = (1.20 \pm 0.12) \times 10^{-7}$ and $L = (1.4 \pm 0.4) \times 10^{-11}\,\rm cm^{-1}$
($\chi^2 = 2.2$ for 10 degrees of freedom); $B$ and $D$ agree within $1.5\,\sigma$ with the laboratory values ($B_0 = 13.9314$ and $D_0 = 1.383 \times 10^{-3}$\,cm$^{-1}$, \citealt{Domenech2018};
$D_0 = 1.376 \times 10^{-3}$\,cm$^{-1}$ from levels $J \le 17$, \citealt{Hakalla2006})\mg{, which confirms that the lines arise in $v = 0$ ($B_1 = 13.44$\,cm$^{-1}$; \citealt{Pearce2024})}. $H$ agrees with $H_0 = 1.10 \times 10^{-7}$\,cm$^{-1}$ \citep{Hakalla2006}
but is about half the $H_0 = 2.2 \times 10^{-7}$\,cm$^{-1}$ of \citet{Domenech2018}, which was constrained by the high-$J$ lines \mg{\citep{Cernicharo1997}}
in a fit without an $L$ term, so the measured lines are consistent with the low-$J$ \mg{$v = 0$} laboratory data once a higher-order
distortion term is included.

\subsection{The radiative cascade and the nascent distribution} \label{sec:cascade}

\rd{In the X\,$^1\Sigma^+$ ground state, which has no electronic angular momentum, rotational transitions obey $\Delta J = \pm 1$, so} the only radiative decay of level $J$ is the R($J{-}1$) transition to $J-1$. With $A = 40$--180\,s$^{-1}$,
the radiative lifetimes of the observed levels are 6--26\,ms, and collisional excitation or de-excitation is negligible
unless the density exceeds $A/k \sim 4 \times 10^{10}$--$2 \times 10^{11}\,\rm cm^{-3}$ \rv{for H$_2$ and $4 \times 10^{11}$--$2 \times 10^{12}\,\rm cm^{-3}$ for H}, where $k \sim 10^{-9}$ \rv{and $10^{-10}$}$\,\rm cm^3\,s^{-1}$ \rv{are} the rate
coefficient\rv{s} for rotational transitions in collisions with \rv{H$_2$ (of order the Langevin rate) and H \citep{Faure2017,GonzalezLezana2026}, respectively}\mg{; such densities are several orders of magnitude larger than is plausible for the molecular gas in HH 168}. Electron collisions are faster for a polar ion
($k_e \approx 3 \times 10^{-7}\,\rm cm^3\,s^{-1}$ for $J = 1 \to 0$ at 2000--3000\,K, from calculations that extend to $J = 11$; \citealt{Hamilton2016,Faure2017}), but the
corresponding critical electron densities, $10^{8}$--$10^{9}\,\rm cm^{-3}$, far exceed \bl{any plausible electron density in the shocked gas}. The main destruction route of CH$^+$, the reverse
of reaction (\ref{eq:form}) with H atoms, has a rate coefficient of $(4$--$8) \times 10^{-10}\,\rm cm^3\,s^{-1}$ at 1000--3000\,K \citep{Faure2017} and
is likewise slow compared with the radiative decay for $n({\rm H}) \ll 10^{10}\,\rm cm^{-3}$.

Each CH$^+$ ion formed
in level $J_{\rm f}$ therefore emits exactly one photon in every R($J$) line with $J < J_{\rm f}$ on its way to the ground
state. The photon intensity of the R($J{-}1$) line, $I_J = F_J / (h\nu_J \Omega)$, is thus the rate per unit area and
solid angle at which CH$^+$ is formed in levels $\geq J$,
\begin{equation}
I_J = \frac{1}{4\pi} \sum_{J' \geq J} P(J'),
\label{eq:cascade}
\end{equation}
where $P(J')$ is the formation rate per unit area into level $J'$, and the nascent distribution follows from
differences of successive lines, $P(J) = 4\pi (I_J - I_{J+1})$. No excitation model is required.

\begin{figure*}[t]
\centering\includegraphics[width=\textwidth,height=0.65\textheight,keepaspectratio]{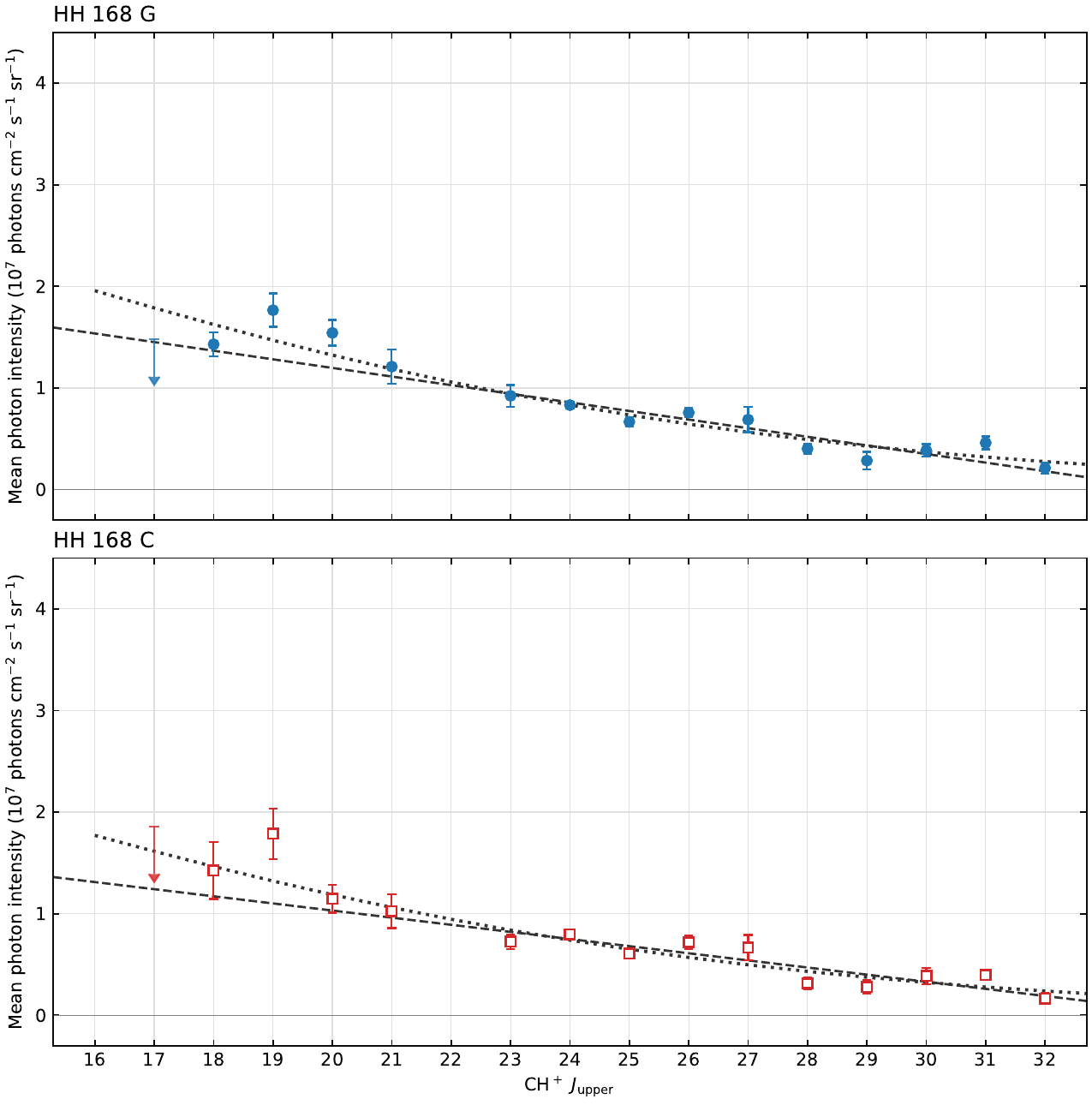}
\caption{Mean photon intensity of the CH$^+$ R($J{-}1$) lines against the upper level $J$, in fields G (top) and C
(bottom), corrected for extinction. Points: lines detected at more than $2.5\,\sigma$ (the detections shown in Figure~\ref{fig:lines} are those above $5\,\sigma$; the fits use all measured intensities with their errors), with $1\,\sigma$ error bars;
arrows: $3\,\sigma$ upper limits. R(21) ($J = 22$), which lies under [Fe\,{\sc ii}] 17.936\,$\mu$m, and R(15) ($J = 16$), blended with an unrelated line, are omitted. By
equation~(\ref{eq:cascade}) each point is the rate at which CH$^+$ is formed in levels $\ge J$. Dashed line: the cascade expected if CH$^+$
were formed at the same rate $p$ into every level up to $J_{\rm max}$ (with no statistical-weight factor)\rv{, so that $I_J = p\,(J_{\rm max} - J + 1)$}, fitted to all
measured values including the non-detections: $J_{\rm max} = 33.1 \pm 0.5$, $p$ $= (0.85 \pm 0.05) \times 10^6$ photons
cm$^{-2}$ s$^{-1}$ sr$^{-1}$ per level, $\chi^2 = \bl{47}$ for 13 degrees of freedom (G); $J_{\rm max} = 33.7 \pm 0.6$,
$p$ $= (0.70 \pm 0.06) \times 10^6$, $\chi^2 = 38$ (C), with statistical errors only; with a 10\% systematic uncertainty added to each line, $\chi^2 = 24$ for 13 degrees of freedom in both fields. \bl{Dotted curve: the cascade from a Boltzmann nascent distribution at the formation temperature $T_{\rm f} = \phi T$, fitted in the same way: $T_{\rm f} = 6400$\,K, $\chi^2 = 35$ (G) and 6300\,K, $\chi^2 = 30$ (C) with statistical errors only, and $T_{\rm f} = 6300$ and 6100\,K, $\chi^2 = 19$ for 13 degrees of freedom in both fields, with the 10\% systematic uncertainty.}
\label{fig:phot}}
\end{figure*}

Figure~\ref{fig:phot} shows $I_J$ for the two fields, corrected for extinction. It falls almost linearly with \mg{increasing} $J$, from $(\bl{1.1}$--$1.8) \times 10^7$ photons cm$^{-2}$ s$^{-1}$
sr$^{-1}$ at $J = 18$--20 to $(1.6$--$2.1) \times 10^6$ at $J = 32$; the two fields agree in shape and within 15\% in normalization. By equation~(\ref{eq:cascade}), a linear decline is expected if CH$^+$ is formed at a similar rate into every level up to some $J_{\rm max}$ \mg{and then cascades radiatively ($\Delta J = -1$) to lower $J$}, for then $I_J \propto J_{\rm max} - J + 1$. A fit of this form (Figure~\ref{fig:phot}, dashed) gives $J_{\rm max} = 33.1 \pm 0.5$ (G) and $33.7 \pm 0.6$ (C) and matches \bl{most lines to within 30\%}; $\chi^2/{\rm dof}$ falls from \bl{3.6} and 2.9 to \bl{1.9} and 1.9 when a 10\% systematic uncertainty is added to each line.

The nascent distribution is therefore broad, with roughly uniform formation, $(0.70$--$0.85) \times 10^6$ photons cm$^{-2}$ s$^{-1}$ sr$^{-1}$ per level, from $J \lesssim 17$ to $J \approx 33$; the largest departure is an excess of \bl{30--60\% at $J = 19$, in R(18).} Formation into levels below $J = 17$, whose lines lie beyond the range used here,
is not constrained\footnote{\mg{The lines R(1)--R(12), at $\approx 28.5$--180\,$\mu$m, which would complete the nascent distribution, could be measured with a high-resolution far-infrared spectrometer such as that proposed for PRIMA \citep{Bradford2025}.}}, and the $J = 17$ line is \bl{only tentative ($\approx 2\,\sigma$)}. The total formation rate into $J \ge 18$ is $4\pi I_{18} \mg{\approx} 1.8 \times 10^8$
cm$^{-2}$ s$^{-1}$ in both fields. A rotation diagram of the same lines, which is the conventional presentation, gives a ``rotational
temperature'' of $\approx 3300$\,K; \mg{this temperature} has no physical meaning here, since the populations are not set by collisions.

\bl{A physically motivated alternative is a thermal nascent distribution. If the energy available above the reaction threshold is distributed roughly exponentially, as it is for reactants with a Maxwellian distribution of collision energies at a temperature $T$, and a fixed fraction $\phi$ of it appears in product rotation, the new ions follow a Boltzmann distribution, $P(J) \propto (2J+1)\exp(-E_J/kT_{\rm f})$, at a formation temperature $T_{\rm f} = \phi T$; only the product $\phi T$ is constrained. Equation~(\ref{eq:cascade}) with this distribution fits the data \gn{slightly} better than uniform formation \gn{($\chi^2$ lower by 5--6 in each field for the same number of free parameters; an improvement this large would arise by chance in $\approx 1$\% of cases if formation were uniform)} (Figure~\ref{fig:phot}, dotted): $T_{\rm f} = 6300 \pm 550$\,K (G) and $6100 \pm 650$\,K (C), with $\chi^2/{\rm dof} = 1.4$ in both \mg{G and C} when \mg{a} 10\% systematic uncertainty is included (2.7 and 2.3 with statistical errors alone). The mean rotational energy of the new ions is then $\approx 0.57$\,eV. Even if all the available energy appeared in rotation, this would require reactant energies equivalent to $T \gtrsim 6000$\,K; for $\phi \approx 2/3$, as quantum calculations suggest at lower energies \citep{Faure2017}, $T \approx 9300$\,K. This distribution would place about half of the formation in levels below $J = 17$, which our data do not constrain.} \rv{Both descriptions rest on simplified and idealized assumptions: uniform formation is an empirical description, adopted because it fits the data rather than derived from the reaction dynamics, and the Boltzmann form assumes a Maxwellian distribution of collision energies with a fixed fraction $\phi$ of the available energy appearing in rotation. A quantitative understanding will require state-to-state calculations of reaction~(\ref{eq:form}) at collision energies above 1\,eV, extending those of \citet[see Section~\ref{sec:discussion}]{GonzalezLezana2026}.}

At the energies involved, some CH$^+$ may also be formed vibrationally excited\rv{, or in the lowest excited electronic state, a\,$^3\Pi$, about 1.2\,eV above the ground state \citep{Amitay1996}}. In $v = 1$, the rotational decay rates within the
vibrational level (54--140\,s$^{-1}$ for $J = 20$--30) exceed the vibrational decay rates to $v = 0$ (9--15\,s$^{-1}$;
\citealt{Pearce2024}), so an ion formed in $v = 1$ emits a rotational line of $v = 1$ at most steps, but has a 10--15\% chance at
each step of decaying to $v = 0$ with $\Delta J = \pm 1$, thereby entering the $v = 0$ ladder at a similar $J$. \rv{Ions formed in a\,$^3\Pi$, which has no spin-allowed decay to the ground state, survive for several seconds \citep{Amitay1996}, still short compared with their destruction time, before decaying to X\,$^1\Sigma^+$; they too can enter the $v = 0$ ladder only indirectly.} The $v = 0$ cascade
therefore measures the rate at which ions enter the $v = 0$ ladder at each $J$, directly or through such decays, rather than the
full nascent distribution. The $v = 1$ rotational lines themselves are not detected: stacked at the positions predicted from the
ExoMol $v = 1$ levels with the offset law of the $v = 0$ lines, they give $2.5\,\sigma$ (G) and $2.0\,\sigma$ (C), and a flux ratio
$F(v{=}1)/F(v{=}0) < 0.31$ in both fields ($3\,\sigma$), so that most of the observed emission comes from ions formed in $v = 0$. \mg{This limit applies only to the high-$J$ levels that MIRI observes. Formation into $v \ge 1$ at low $J$ remains possible: reaction~(\ref{eq:form}) with vibrationally excited H$_2$ is exothermic, and much of the reactant vibrational energy may appear as product vibration, just as H$_2$O photodissociated in its first absorption band yields OH that is vibrationally hot but rotationally cold \citep{Tabone2021}. Ions formed in this way would emit only far-infrared rotational lines and the $v = 1 \to 0$ band at 3.5--4.1\,$\mu$m, observed in NGC 7027 \citep{Neufeld2021}, so the $v = 0$ cascade gives a lower limit to the total formation rate.}

The highest level detected, $J = 32$ at $E/k = 19{,}100$\,K, places a lower limit on the energy available in
reaction (\ref{eq:form}). Since the products must carry away the endothermicity as well as the rotational energy, the
reactants must supply at least $1.65 + 0.40 = 2.05$\,eV. This is the sum of the center-of-mass collision energy and the internal
energy of the H$_2$, which can be substantial; \rv{the reaction is exothermic for vibrationally excited H$_2$, and also for H$_2$ in $v = 0$ with $J \geq 8$ \citep{Faure2017}}. If the H$_2$ were in low
levels, the collision energy alone would need to reach 2.05\,eV, which for ions drifting through H$_2$ in a C-type shock requires a
drift speed of at least 15\,\kms\ (reduced mass 1.73\,amu), and for thermal collisions requires the tail of the Maxwellian beyond
$\sim 2$\,eV, i.e., $T \gtrsim 7000$\,K; reactions with excited H$_2$ relax these limits. \rv{For comparison, the H$_2$ rotational excitation temperatures in the two fields rise from 700--900\,K (S(1)--S(3)) to 1100--1500\,K (S(5)--S(8)); these approximate the kinetic temperatures of the H$_2$-emitting gas if its levels are close to equilibrium, and at such temperatures only a few per cent of the H$_2$ is in levels with $J \geq 8$. These temperatures, far below 7000\,K, are typical of C-type shocks, in which the energy for reaction~(\ref{eq:form}) comes from ion-neutral drift rather than from the gas temperature. They would also be consistent with a J-type shock if the gas spans a range of temperatures, since the H$_2$ emission would then come mostly from gas cooler than that in which the highest CH$^+$ levels form. The drift speed cannot be measured directly, because the MIRI line profiles do not resolve it, but the required $\gtrsim 15$\,\kms\ is plausible in C-type shocks, in which the ion-neutral drift speed can be a substantial fraction of the shock speed \citep{Draine1980}.} The lines detected at
$J = 29$--32 follow the extrapolation of the $J = 21$--28 decline to within 0.3\,dex, and no clear turnover is seen, although these lines are detected at only 3--8\,$\sigma$: the distribution
may well extend beyond the highest level we can measure.

\section{Discussion} \label{sec:discussion}

The high-$J$ CH$^+$ lines open an unusual observational window. In laboratory ion-beam
experiments the product state distribution of reaction (\ref{eq:form}) has been measured only for the electronically
excited A\,$^1\Pi$ product, through its chemiluminescence \citep{Kusunoki1979}; the ground-state product distribution, which determines the
chemical pumping invoked to explain CH$^+$ excitation throughout \gn{most of} the interstellar medium \citetext{\citealt{Godard2013,Faure2017}\gn{; but see \citealt{Oka2013} for radiative pumping toward Herschel 36}},
has been accessible only through theory. The HH 168 spectrum provides it directly, albeit averaged over the
collision energies of the shock: formation at a roughly uniform rate per level from $J \lesssim 17$ to $J \approx 33$\bl{, or a Boltzmann distribution at $\phi T \approx 6200$\,K}. The situation parallels that of the suprathermal OH lines, whose
intensities map the rotational distribution of OH emerging from H$_2$O photodissociation \citep{Tappe2008,Tabone2021,Zannese2024,Neufeld2024}; here the
process is a bimolecular reaction rather than a photodissociation. A quantum \rv{or statistical} calculation of reaction (\ref{eq:form}) on the
ground-state CH$_2^+$ surface at collision energies of 1--3\,eV, such as those performed by \citet{Zanchet2013} \rv{and \citet{GonzalezLezana2026}} at lower
energies, would be tested by these data.

\bl{The lines also constrain the CH$^+$ budget of the shock. CH$^+$ is destroyed rapidly by reactions with H, H$_2$ and electrons, so the line intensities measure the rate of reaction (\ref{eq:form}) rather than the abundance of CH$^+$: the observed formation rate into $J \geq 18$, $1.8 \times 10^8$ ions cm$^{-2}$ s$^{-1}$, is the product of the C$^+$ abundance, the density of reactive H$_2$ and the rate coefficient, integrated through the emitting layer. Carbon nuclei enter the shock at a rate $x_{\rm C}\, n_{\rm H}\, v_s = 10^8\, x_{-4}\, n_5\, v_7$ cm$^{-2}$ s$^{-1}$, where $x_{\rm C} = 10^{-4}\, x_{-4}$ is the gas-phase carbon abundance per H nucleus, $n_{\rm H} = 10^5\, n_5\,\rm cm^{-3}$ the preshock density and $v_s = 100\, v_7\,\rm km\,s^{-1}$ the shock speed. For a shock surface viewed face-on, each carbon nucleus must therefore form $\approx 1.8/(x_{-4}\, n_5\, v_7)$ CH$^+$ ions in levels $J \geq 18$, or $\approx 5/(x_{-4}\, n_5\, v_7)$ in all levels if the Boltzmann distribution is extrapolated to low $J$. \rd{The observations therefore require $N_{\rm cyc}\, x_{-4}\, n_5\, v_7 \gtrsim 2$--5, where $N_{\rm cyc}$ is the number of times each carbon nucleus forms CH$^+$. Because the reaction of CH$^+$ with H returns the carbon to C$^+$ within the hot layer, $N_{\rm cyc} > 1$ is expected, and the data do not distinguish a large preshock mass flux from repeated cycling.} Models of C- and J-type shocks that follow the formation, excitation and destruction of CH$^+$ will be presented in a future paper.}

Finally, these lines should be present wherever H$_2$ is heated to a few thousand kelvin with little infrared radiation
to confuse the spectrum. \rvb{These lines are detected in at least one other protostellar shock observed with MIRI, associated with HOPS 78, as we will report separately (S.~Lee, PhD thesis in preparation; S.~Lee et al., ApJ, in preparation).} \bl{In HH 168, the} lines are not blended with any strong feature in
the 13.7--21.3\,$\mu$m range except R(21); R(19), R(26), R(28), R(29) and R(31) are fitted with neighboring lines as extra components. Where the formation energy comes from ion-neutral drift rather than thermal
collisions, the shape of the cascade should differ, and the two can be told apart.

\section{Conclusions} \label{sec:conclusions}

\begin{enumerate}
\item We have detected the pure rotational lines \bl{R(17)}--R(31) of CH$^+$, from levels $J = \bl{18}$--32 up to 19,100\,K above the
ground state, toward HH 168 with JWST/MIRI; twelve lines in one field and eleven in the other exceed 5\,$\sigma$, and their velocities and
maps match those of the hot H$_2$.
\item The measured line positions are up to 150\,\kms\ redward of the ExoMol predictions, the offset increasing with $J$,
and constitute the first measurement of the CH$^+$ rotational intervals above $J = 16$.
\item Because the lines are the radiative cascade of newly formed ions, their photon intensities give the nascent
rotational distribution of the C$^+$ + H$_2$ reaction products entering the $v = 0$ ladder: roughly uniform formation per level from $J \lesssim 17$ (lower levels are not constrained) to $J \approx 33$, which requires at
least 2\,eV of reactant energy. The formation
rate into $J \geq 18$ is $1.8 \times 10^8$ ions cm$^{-2}$ s$^{-1}$.
\item \bl{The cascade is fitted \gn{slightly} better \gn{($\chi^2$ lower by 5--6 in each field, for the same number of free parameters)} by a Boltzmann nascent distribution at a formation temperature $\phi T = 6100$--6300\,K, where $\phi$ is the fraction of the available energy that appears in product rotation, than by uniform formation per level\gn{, although neither is excluded}. This requires reactant energies equivalent to $T \gtrsim 6000$\,K, or $\approx 9300$\,K if $\phi \approx 2/3$.}
\end{enumerate}

\begin{acknowledgments}
This work is based on observations made with the NASA/ESA/CSA James Webb Space Telescope. The data were obtained from the
Mikulski Archive for Space Telescopes at the Space Telescope Science Institute, which is operated by the Association of
Universities for Research in Astronomy, Inc., under NASA contract NAS 5-03127 for JWST. These observations are associated
with program \#7084. Support for program \#7084 was provided by NASA through grant JWST-GO-07084 from the Space Telescope
Science Institute. D.A.N.\ was supported in part by JWST Theory grant JWST-AR-06829, ``Understanding the rich molecular spectra
observed with NIRSpec and MIRI/MRS toward protostellar outflows.''

Use of generative AI: Anthropic's Claude (Opus 5.5 and Fable 5.1, accessed through Claude Code) was used for data analysis, coding, generating the figures from the data using Matplotlib, and drafting and revising the manuscript text. OpenAI's Codex was used to write the initial line-fitting and mapping code. The authors take full responsibility for the content of this paper.
\end{acknowledgments}

\facility{JWST (MIRI)}

\software{Matplotlib \citep{Hunter2007}}

\bibliographystyle{aasjournal}
\bibliography{refs}

\clearpage

\end{document}